\documentclass[12pt]{article}

\begin{document}

\begin{titlepage}
\vspace*{5mm}
\begin{center} {\Large \bf $p$--species integrable reaction--diffusion
processes }\\ \vskip 1cm
{\bf M. Alimohammadi$^{a,b}$ \footnote {e-mail:alimohmd@theory.ipm.ac.ir},
N. Ahmadi$^a$}\\

\vskip 1cm
{\it $^a$ Physics Department, University of Tehran, North Karegar Ave.,} \\
{\it Tehran, Iran }\\
{\it $^b$ Institute for Studies in Theoretical Physics and Mathematics,}\\
{\it  P.O.Box 5531, Tehran 19395, Iran}
\end{center}
\vskip 2cm
\begin{abstract}
We consider a process in which there are $p$--species of particles, i.e. $%
A_1,A_2,\cdots ,A_p$, on an infinite one--dimensional lattice. Each particle
$A_i$
can diffuse to its right neighboring site with rate $D_i$, if this site is
not already occupied. Also they have the exchange interaction $%
A_j+A_i\rightarrow A_i+A_j$ with rate $r_{ij}.$ We study the range of
parameters (interactions) for which the model is integrable. The
wavefunctions of this multi--parameter family of integrable models are
found. We also extend
the 2--species model to the case in which the particles are able to diffuse to
their right or left neighboring sites.
\end{abstract}

\begin{center}
 \\
\end{center}
\end{titlepage}
\newpage
\section{ Introduction}

Our understanding of nonequilibrium statistical physics is far
behind that for the equilibrium theory. Even simple models may
pose a formidable problem if one wants to approach them
analytically. As an interesting example of stochastic models,
which may be investigated analytically in some few cases, are
one--dimensional reaction--diffusion processes which are of both
theoretical and experimental interest in a very wide context of
physics and chemistry, such as stochastic spin flip dynamics
\cite{r1}, traffic flow \cite{r2,r3}, the kinetics of
bipolymerization \cite{r4,r5}, reptation of DNA in gels
\cite{r6,r7}, interface growth \cite{r8,r9}, diffusion in zeolites
\cite{r10,r11}, and many other phenomena.

Asymmetric Simple Exclusion Processes (ASEP) in one dimension, is one of the
simplest example of a driven diffusion system \cite{r12,r13}. For example,
the totally ASEP model describes a process in which each lattice site can be
occupied by at most one particle and the particles hop to their right
neighboring site if they are not already occupied, with a rate which is the
same for all particles, otherwise the attempted move is rejected. The
dynamics of these models can be fully specified by a master equation and an
appropriate boundary condition, which imposed on the probabilities appear in
the master equation. Using the coordinate Bethe ansatz, the author of \cite
{r14} has exactly obtained the $N$--particle conditional probabilities of
totally ASEP, in which the particles can move to left and right with
different rates.

Now the interesting point is that if one changes the boundary condition,
without altering the master equation, one can model another
reaction--diffusion processes even with long range interactions. For example
in \cite{r15}, the so called ''generalized totally ASEP model'' has been
exactly solved in this way. In this model the particle hops to the next
right site by pushing all the neighboring particles to their next right
sites, with a rate depending on the number of right neighboring particles.
The partially generalized ASEP model has also been studied in \cite{r16}.
Note that in all of these cases, the solvability of the models is
shown by proving the factorization of $N$--particle S--matrices into
2--particle ones, which the latter were found exactly.

In all the above ASEPs, there is only one species of particle,
that is all the particles are of the same type. But if one
considers the two, or more, species problems, the situation
becomes more complicated. The main complexity arises from the fact
that the above mentioned factorization of $N$--particle systems,
reduces to the condition of satisfying the two--particle
S--matrices in the Quantum Yang--Baxter Equation (QYBE). In one
species models, the S--matrices are not really matrices, they are
c--numbers and therefore their satisfying in QYBE becomes trivial.
This new condition can hardly restrict the number of solvable
models with more than one--species particles. In \cite{r17}, a
class of two species reaction--diffusion processes with following
properties has been considered: 1) the particles diffuse to their
right neighboring sites, 2) they can be annihilated or created,
but the total number of particles are constant, and 3) the
interaction rates are all the same. It is shown that among 4096
types of models with the above properties, which can be modeled by
a master equation and a number of boundary conditions, there are
only 28 independent interactions which their two--particle
S--matrices satisfy the QYBE and therefore are solvable. The third
condition (equality of the interaction rates) was very crucial in
proof of solvability.

In this paper we want to study the effect of interaction rates in
solvability of $p$--species reaction--diffusion processes, by
considering a specific model. We begin our investigation by
choosing one of the two--species interactions that has been
introduced in \cite{r17}, but with different interaction rates, and
try to obtain the range of parameters to insure the solvability of
the corresponding extended $p$--species model. As we show, we must
restrict ourselves to a narrower and narrower range of parameters,
as we go ahead, and finally arrive at a model with a specific
relation between the interaction rates and also a specific range
for these rates.

The plan of the paper is as following. In section 2, we begin with following
two--species reaction--diffusion processes:
$$ A+\emptyset \stackrel{D_A}{\rightarrow }\emptyset +A, 
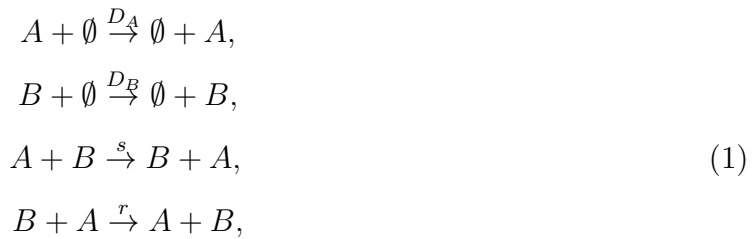
$$
$$ B+\emptyset \stackrel{D_B}{\rightarrow }\emptyset +B,$$
\begin{equation}
A+B\stackrel{s}{\rightarrow }B+A,
\label{1}
\end{equation}
$$ B+A\stackrel{r}{\rightarrow }A+B,$$
and write down a master equation and a number of boundary conditions to
describe the dynamics of these interactions. $D_A$ and $D_B$ are the
right--diffusion rates of $A$ and $B$ particles, respectively, and $s$ and $r$
are the rates of transforming (exchanging) $A$ and $B$ particles to each
other for (...$AB$...) and (...$BA$...) configurations, respectively. We will
show that only for $D_A=D_B=1$ case, there exists the coordinate
Bethe--ansatz solution for probabilities (note that taking $D\equiv 1$ is in
fact a choosing of time scale). Moreover, we will see that the consistency
of the solutions (which will be appeared as satisfaction of 2--particle
S--matrix in QYBE) restricts us to $r=0$ or $s=0$ cases (which are the same
after relabeling $A\longleftrightarrow B$). We therefore conclude that the
solvable model (interaction) is the following one--parameter family process
$$ A+\emptyset \stackrel{1}{\rightarrow }\emptyset +A, 
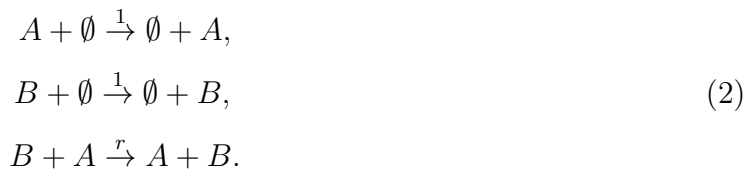
$$
\begin{equation}
B+\emptyset \stackrel{1}{\rightarrow }\emptyset +B,
\label{2}
\end{equation}
$$ B+A\stackrel{r}{\rightarrow }A+B. $$
Note that for $r=1$, interactions (\ref{2}) are one of the 28
interactions introduced in \cite{r17}. We then generalize the
interactions (\ref{2}) to a $p$--species model in which the particles
$A_i(i=1,...,p)$ can diffuse to their right neighboring sites, all
with equal rate one, and also they have exchange interactions with different
rates:
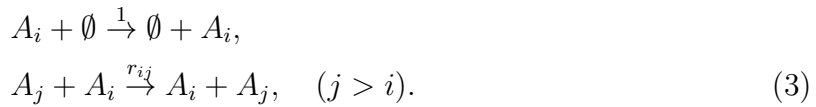
\begin{eqnarray}
&&A_i+\emptyset \stackrel{1}{\rightarrow }\emptyset +A_i,\nonumber \\
&&A_j+A_i\stackrel{}{\stackrel{r_{ij}}{\rightarrow }A_i+A_j,\quad (j>i).}
\label{3}
\end{eqnarray}
We label the species such that the configuration (...$A_jA_i$...) can go to
(...$A_iA_j$...), only when $j>i$, therefore $r$ in eq.(\ref{2}) is in fact $%
r_{12}$. After a lengthy calculation, we show that there must be a specific
relations between $r_{ij}$s until the $p^2\times p^2$ two--particle
S--matrix satisfying the QYBE.

In section 3, we calculate the two--particle conditional probabilities of
reaction (\ref{2}), and show that only for $0\leq r<2$ we are able to
calculate these probabilities by a standard superposition of the
eigenfunctions with real eigenvalues. The long--time behavior of the
probabilities is also discussed. Finally in section 4, we generalize the
reactions (\ref{2}) to the case where the particles can diffuse to both
right and left, as following:
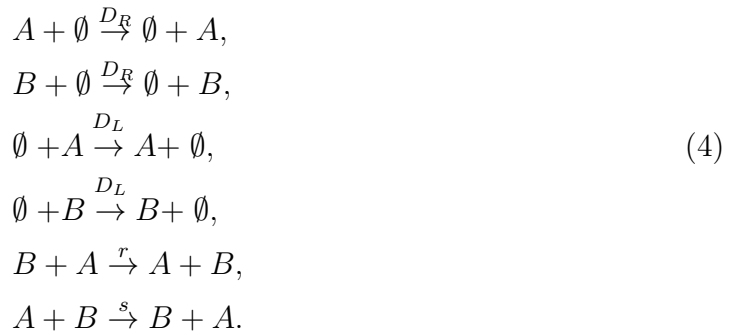
\begin{eqnarray}
&&A+\emptyset \stackrel{D_R}{\rightarrow }\emptyset +A, \nonumber\\
&&B+\emptyset \stackrel{D_R}{\rightarrow }\emptyset +B,  \nonumber \\
&&\emptyset \stackrel{D_L}{+A\rightarrow A+}\emptyset ,  \\
&&\emptyset \stackrel{D_L}{+B\rightarrow B+}\emptyset ,  \nonumber \\
&&B+A\stackrel{r}{\rightarrow }A+B,  \nonumber \\
&&A+B\stackrel{s}{\rightarrow }B+A.  \nonumber
\label{4}
\end{eqnarray}
We show that there must be a fine tuning of parameters if one demands the
reactions (4) to be solvable.

\section{ $p$--species exchange--diffusion processes}

\subsection{ The master equation for 2--species case}

Consider the interactions introduced in eq.(\ref{1}). The basic
quantities that must be calculated are the probabilities
$P_{\alpha _1\alpha _2...\alpha _N}(x_1,x_2,...,x_N;t)$ for
finding at time $t$ the particle of type $\alpha _1$ at site
$x_1$, particle of type $\alpha _2$ at site $x_2$, etc.. Each
$\alpha _i$ can be $A$ or $B$. Following \cite{r14}, we take these
functions to define probabilities only in the physical region $%
x_1<x_2<...<x_N$, and the regions where any two adjacent
coordinates are
equal, are the boundaries of the physical region. For $x_{i-1}-x_i>1,\forall
i$, the particles can only hop to their right neighboring sites and
therefore the master equation is:
$$ \frac \partial {\partial t}P_{\alpha _1\alpha _2...\alpha
_N}(x_1,x_2,...,x_N;t)=\sum\limits_{i=1}^ND_{\alpha _i}[ P_{\alpha
_1\alpha _2...\alpha _N}(x_1,...,x_{i-1},x_i-1,x_{i+1},...,x_N;t)$$
\begin{equation}
-P_{\alpha_1\alpha _2...\alpha _N}(x_1,x_2,...,x_N;t)] , \label{5}
\end{equation}
where the first $N$ terms are the sources of $P_{\alpha _1\alpha _2...\alpha
_N}(x_1,x_2,...,x_N;t)$ and the second $N$ terms are the sinks of it. It is
obvious that if $x_{i+1}=x_i+1$ for some $i$'s, then some of the probability
functions in the right hand side of eq.(\ref{5}) go out from the physical
region. So we need to specify the boundary terms. The specification of these
terms depends on the details of the interactions of the particles. For
exchange interactions defined in eq.(\ref{1}), the suitable boundary
conditions are:
\begin{eqnarray}
D_AP_{BA}(x,x) &=&sP_{AB}(x,x+1)+(D_B-r)P_{BA}(x,x+1), \nonumber \\
D_BP_{AB}(x,x) &=&rP_{BA}(x,x+1)+(D_A-s)P_{AB}(x,x+1),  \\
P_{\alpha \alpha }(x,x) &=&P_{\alpha \alpha }(x,x+1)~,\qquad
(\alpha =A,B), \nonumber
\end{eqnarray}
in which the time variable and all the other coordinates have been
suppressed for simplicity. To justify these boundary conditions, it is
enough to examine them in some specific cases. Let us do it for a rather
complicated case, for example $P_{ABBA}(x,x+1,x+2,x+3).$ From master
equation (\ref{5}), we have,
$$
\frac \partial {\partial t}P_{ABBA}(x,x+1,x+2,x+3)
=D_AP_{ABBA}(x-1,x+1,x+2,x+3)$$
\begin{equation}
+D_BP_{ABBA}(x,x,x+2,x+3)+D_BP_{ABBA}(x,x+1,x+1,x+3)
\end{equation}
$$+D_AP_{ABBA}(x,x+1,x+2,x+2)-2(D_A+D_B)P_{ABBA}(x,x+1,x+2,x+3).$$
If we use the relations (6) in the second, third and fourth terms of
the right hand side of (7), we find
$$
\frac \partial {\partial t}P_{ABBA}(x,x+1,x+2,x+3)
=D_AP_{ABBA}(x-1,x+1,x+2,x+3)+$$
\begin{equation}
rP_{BABA}(x,x+1,x+2,x+3)+sP_{ABAB}(x,x+1,x+2,x+3)
-(D_A+r+s)P_{ABBA}(x,x+1,x+2,x+3).  \nonumber
\end{equation}
This equation is exactly what we expect from interactions (\ref{1}), because
the source terms of configuration (...$\emptyset ABBA\emptyset $...) are:
(...$A\emptyset BBA\emptyset $...) (with rate $D_A$), (...$\emptyset
BABA\emptyset $...) (with rate $r$) and (...$\emptyset ABAB\emptyset $...)
(with rate $s$), and its sink terms are: (...$\emptyset ABB\emptyset A$...)
(with rate $D_A$), (...$\emptyset BABA\emptyset $...) (with rate $s$) and
(...$\emptyset ABAB\emptyset $...) (with rate $r$). It can be shown that the
boundary conditions (6) results the correct terms for any desired
configuration.

\subsection{ the Bethe ansatz solution (2--species)}

Now we want to solve the master equation (5) with boundary conditions
(6) by the coordinate Bethe ansatz method. First we define $\Psi
_{\alpha _1...\alpha _N}(x_1,...,x_N)$ through,
\begin{equation}
P_{\alpha _1...\alpha _N}(x_1,...,x_N;t)=e^{-\epsilon _Nt}\Psi _{\alpha
_1...\alpha _N}(x_1,...,x_N),
\end{equation}
and then substitute it in eq.(5), which results
\begin{equation}
\sum\limits_{i=1}^ND_{\alpha _i}\Psi _{\alpha _1...\alpha
_N}(x_1,...,x_{i-1},x_i-1,x_{i+1},...,x_N)=(\sum\limits_{i=1}^ND_{\alpha
_i}-\epsilon _N)\Psi _{\alpha _1...\alpha _N}(x_1,...,x_N).
\end{equation}
To solve this equation, we use the coordinate Bethe ansatz for each of the
components $\Psi _{\alpha _1...\alpha _N}(x_1,...,x_N):$%
\begin{equation}
\Psi _{\alpha _1...\alpha _N}(x_1,...,x_N)=\sum\limits_\sigma ^{}A_\sigma
^{(\alpha _1...\alpha _N)}e^{i\sigma (\mathbf{p}).\mathbf{x}},
\end{equation}
where $\mathbf{x}$ and $\mathbf{p}$ stands for $N$--tuples coordinates and
momenta, respectively, and $\sigma (\mathbf{p})$ is a permutation of
momenta. The sum is over all permutations. Inserting (11) into
(10) yields:
\begin{equation}
\sum\limits_\sigma ^{}\left[ \epsilon _N-\sum\limits_{j=1}^ND_{\alpha
_j}+\sum\limits_{j=1}^ND_{\alpha _j}e^{-i\sigma
(p_j)}\right] A_\sigma ^{(\alpha _1...\alpha _N)}e^{i\sigma (\mathbf{p}).%
\mathbf{x}}=0.
\end{equation}
As $A_\sigma ^{(\alpha _1...\alpha _N)}e^{i\sigma (\mathbf{p}).\mathbf{x}}$s
are linearly independent for different $\sigma ^{\prime }$s, the only
solution of eq.(12) is:
\begin{equation}
\epsilon _N-\sum\limits_{j=1}^ND_{\alpha
_j}+\sum\limits_{j=1}^ND_{\alpha _j}e^{-i\sigma (p_j)}=0,\ \
\forall \sigma ,
\end{equation}
or
\begin{equation}
\sum\limits_{j=1}^ND_{\alpha _j}e^{-i\sigma
_1(p_j)}=\sum\limits_{j=1}^ND_{\alpha _j}e^{-i\sigma
_2(p_j)}=...=\sum\limits_{j=1}^ND_{\alpha _j}e^{-i\sigma _n(p_j)},
\end{equation}
where $n$ is the number of elements of permutation group. As these
equalities must hold for an arbitrary $\mathbf{p}$, the only nontrivial
solution is:
\begin{equation}
D_A=D_B\equiv 1.
\end{equation}
Now as for any group element $\sigma $ we have
\begin{equation}
\sum\limits_{j=1}^Ne^{-i\sigma (p_j)}=e^{-ip_1}+...+e^{-ip_N},
\end{equation}
the equalities (14) satisfy satisfactorily. Therefore, the Bethe
ansatz solution exists only for equal diffusion rates, which in this case the
eigenvalue $\epsilon _N$ is found to be (by eq.(13)):
\begin{equation}
\epsilon _N=\sum\limits_{j=1}^N(1-e^{-ip_j}).
\end{equation}
It is easier to consider $\Psi _{\alpha _1...\alpha _N}(x_1,...,x_N)$ and $%
A_\sigma ^{(\alpha _1...\alpha _N)}$ as the components of the tensors $%
\mathbf{\Psi }$ and $\mathbf{A}_\sigma $with rank $N$, respectively.
therefore eq.(11) can be written as
\begin{equation}
\mathbf{\Psi }(x_1,...,x_N)=\sum\limits_\sigma \mathbf{A}_\sigma e^{i\sigma (%
\mathbf{p}).\mathbf{x}}.
\end{equation}
The boundary conditions of $\mathbf{\Psi }$ can be obtained by substituting
eq.(9) in (6), with $D_A=D_B\equiv 1.$ The resulting equation is
\begin{equation}
\mathbf{\Psi }(...,\zeta ,\zeta ,...)=\mathbf{b}_{k,k+1}\mathbf{\Psi }%
(...,\zeta ,\zeta +1,...),
\end{equation}
where
\begin{equation}
\begin{array}{ccc}
{\mathbf{b}}_{k,k+1}={\mathbf{1}}\otimes \cdot \cdot \cdot \otimes\mathbf{1}\otimes
& \underbrace{\mathbf{b}}
& \otimes\mathbf{1}\otimes \cdot \cdot \cdot \otimes {\mathbf{1}}, \\
& k,k+1 &
\end{array}
\label{13}
\end{equation}
with
\begin{equation}
\mathbf{b=}\left(
\begin{array}{cccc}
1 & 0 & 0 & 0 \\
0 & 1-s & r & 0 \\
0 & s & 1-r & 0 \\
0 & 0 & 0 & 1
\end{array}
\right) .
\end{equation}
The coefficients $\mathbf{A}_\sigma $'s in eq.(18) must be found by
substituting the wavefunction (18) into the boundary condition
(19), which yields
\begin{equation}
\sum\limits_\sigma e^{i\sum\limits_{j\neq k,k+1}\sigma (p_j)x_j+i(\sigma
(p_k)+\sigma (p_{k+1}))\xi }(1-e^{i\sigma (p_{k+1})}\mathbf{b}%
_{k,k+1})\mathbf{A}_\sigma =0.
\end{equation}
As the exponential part of eq.(22) is symmetric with respect to $%
p_k\longleftrightarrow p_{k+1}$, if we also symmetrize the remaining terms
with respect to this interchange, we obtain
\begin{equation}
(1-e_{}^{i\sigma (p_{k+1})}\mathbf{b}_{k,k+1})\mathbf{A}_\sigma
+(1-e^{i\sigma (p_k)}\mathbf{b}_{k,k+1})\mathbf{A}_{\sigma \sigma _k}=0,
\end{equation}
where $\sigma _k$ represent the permutation group element which only
interchanges $p_k$ and $p_{k+1}.$ Therefore
\begin{equation}
\mathbf{A}_{\sigma \sigma _k}=\mathbf{S}_{k,k+1}(\sigma (p_k),\sigma
(p_{k+1}))\mathbf{A}_\sigma ,
\end{equation}
where
\begin{equation}
\begin{array}{ccc}
\mathbf{S}_{k,k+1}(z_1,z_2)=\mathbf{1}\otimes \cdot \cdot \cdot \otimes
\mathbf{1}\otimes & \underbrace{S(z_1,z_2)} & \otimes \mathbf{1}\otimes
\cdot \cdot \cdot \otimes \mathbf{1,} \\
& k,k+1 &
\end{array}
\end{equation}
in which
\begin{equation}
\mathbf{S(}z_1,z_2)=-(\mathbf{1-}z_{1\mathbf{\ }}\mathbf{b})^{-1}(\mathbf{1}%
-z_2\mathbf{b).}
\end{equation}
In above equations, $z_k$ stands for $e^{ip_k}$. In this way, all the $%
\mathbf{A}_\sigma $ coefficients are determined in terms of $\mathbf{A}_1$
which is fixed by the initial conditions (the particles' positions at $t=0$%
). It seems that we have solved the problem for arbitrary $\mathbf{b}$ (i.e.
interaction), but it is not true (note that we have not yet used the
explicit form of $\mathbf{b}$ in deriving $\mathbf{A}_\sigma $). The crucial
point is that $\sigma _1\sigma _2\sigma _1$ and $\sigma _2\sigma _1\sigma _2$
are equal as elements of permutation group, therefore we should impose the
following condition on the corresponding $\mathbf{A}_\sigma $'s:
\begin{equation}
\mathbf{A}_{\sigma _1\sigma _2\sigma _1}=\mathbf{A}_{\sigma _2\sigma
_1\sigma _2},
\end{equation}
and this highly restricts the allowed $\mathbf{b}$ matrices (i.e.
interactions ). It can be shown that eq.(27) reduces to following
relation for $S(z_1,z_2)$ matrices (see \cite{r17} for more details),
\begin{equation}
(S(w,t)\otimes 1\mathbf{)(}1\mathbf{\otimes }S\mathbf{(}z,t))(S(z,w)\otimes
1)=(1\otimes S(z,w))(S(z,t)\otimes 1)(1\otimes S(w,t)),
\end{equation}
in which $z=e^{ip_1},w=e^{ip_2},$ and $t=e^{ip_3}$. Note that eq.(28)
is nothing but the Quantum Yang--Baxter equation.

Now if one calculates the S--matrix from eq.(26), using $\mathbf{b}$
from (21), the QYBE (28) reduces to a $8\times 8$ matrix with
fourteen nonzero elements (after writing eq.(28) as RHS -- LHS = 0) that
must be equated to the zero matrix. The elements are functions of $%
p_1,p_2,p_3,r$ and $s$ which all must be equal to zero for arbitrary
momentum values $p_1,p_2$ and $p_3$. It can be shown that the unique
solutions of these fourteen equations are:
\begin{eqnarray}
{\rm solution } \ \ 1 &:& \ \ r=0, \ \ {\rm  arbitrary } \ \ s, \nonumber \\
{\rm solution } \ \ 2 &:& \ \ s=0, \ \ {\rm  arbitrary } \ \ r.
\end{eqnarray}
As these two solutions are equivalent (by relabeling $A\longleftrightarrow B$%
), so the only integrable model is the one indicated in eq.(2).

\subsection{ The $p$--species model}

Now let us generalize the reaction (2) to the case where there exists $%
p$ kinds of particles, which we label them by $A_1,A_2,...,A_p.$ Each
particle can diffuse to its right neighboring site, and any two particle can
exchange with each other. From the results obtained in previous subsection,
we know that if we want the model to be integrable, we must restrict
ourselves to the case where the particles' diffusion rates are equal (scaled
to one), and also for each two particles there is only one allowed exchange
interaction. For example $A_2+A_1\rightarrow A_1+A_2$ is allowed, but $%
A_1+A_2\rightarrow A_2+A_1$ is forbidden. We label the particles such that $%
A_j+A_i\rightarrow A_i+A_j$ is allowed only for $j>i$, and denote the
reaction rate of this interaction by $r_{ij}$, see eq.(3).

The master equation of this $p$--species model is again eq.(5) (with $%
D_{\alpha _i}=1$), but now each $\alpha _i$ can be $A_1,A_2,...,$ and $A_p$.
The boundary conditions are the generalization of eq.(6) but now for each
two particle species $A_i$ and $A_j$, i.e. $P_{AB}\rightarrow P_{ij}$, $%
P_{BA}\rightarrow P_{ji}$, $s=0$, $D_A=D_B=1$, and $r\rightarrow r_{ij}$; so
\begin{eqnarray}
P_{ij}(x,x) &=&P_{ij}(x,x+1)+r_{ij}P_{ji}(x,x+1),\qquad j>i, \nonumber \\
P_{ji}(x,x) &=&(1-r_{ij})P_{ji}(x,x+1),\qquad j>i,   \\
P_{ii}(x,x) &=&P_{ii}(x,x+1).  \nonumber
\end{eqnarray}
The wavefunctions can be again factorized by the eq.(9), and the
coordinate Bethe ansatz solution (18) is yet valid. The boundary
conditions can be rewritten as (19), but here $\mathbf{b}$ is the
following $p^2\times p^2$ matrix:
\begin{equation}
\mathbf{b=}\sum\limits_{i\leq j}^{}E_{ii}\otimes
E_{jj}+\sum\limits_{i<j}^{}r_{ij}E_{ij}\otimes
E_{ji}+\sum\limits_{i>j}^{}(1-r_{ji})E_{ii}\otimes E_{jj},
\end{equation}
where $E_{ij}$ is a $p\times p$ matrix with elements $(E_{ij})_{kl}=%
\delta _{ik}\delta _{jl}$. It can be shown that the S--matrix (26)
becomes:
\begin{equation}
S(z,w)=\sum\limits_{i,j=1}^p\frac{1-w(1-r_{ji}^{\prime })}{(1-r_{ji}^{\prime
})z-1}E_{ii}\otimes E_{jj}+\sum\limits_{i,j=1}^pr_{ij}^{\prime }\frac{z-w}{%
(z-1)(1-z+zr_{ij}^{\prime })}E_{ij}\otimes E_{ji,}
\end{equation}
in which,
\begin{equation}
r_{ij}^{\prime }=\left\{
\begin{array}{c}
0\qquad {\rm if } \ \ i\geq j \\
r_{ij}\qquad {\rm if } \ \ i<j
\end{array}
\right. .
\end{equation}
Now expression (32) must satisfy the QYBE (28). After a lengthy
calculation, it can be shown that the only nontrivial solutions of QYBE are
as following (for each $i<j<k$ indices):
\begin{eqnarray}
{\rm solution } \ \ 1 &:& \ \  r_{ij}=0,\ \ r_{ik} \ {\rm  and } \ r_{jk} \ \ {\rm
arbitrary,} \nonumber \\
{\rm solution }\ \ 2 &:&\ \ r_{ik}=0,\ \ r_{jk}=0, \ {\rm  and } \ r_{ij} \ \ {\rm
arbitrary,}   \\
{\rm solution } \ \ 3 &:&\ \ r_{ij}=r_{ik},\ \  {\rm  }r_{jk} \ \ {\rm
arbitrary.}  \nonumber
\end{eqnarray}
For any set of interaction rates where each three of them satisfy
any of the solutions $1,2$ or $3$, with the constraint that the
relations between $r_{ij}$s must be consistent in all subsets, we have
an integrable $p$--species model with wavefunction (18) whose
coefficients are determined by eq.(24) and S--matrix introduced
in (32).

For $p=3$, the allowed sets of interaction rates are exactly the same as the
three solutions (34) with $(ijk)=(123)$. But for $p>3$ cases, we can
choose different consistent solutions for any $(ijk)$'s and therefore
extracting all the allowed sets are not so easy. For example for $p=4$,
in which there are six interaction rates $r_{12}, r_{13}, r_{14}, r_{23},
r_{24},$ and $r_{34}$, the allowed sets of parameters are as following:
\begin{eqnarray}
&&\left\{ r_{14},r_{24},r_{34}\right\} ,\nonumber \\
&&\left\{ r_{12},r_{34} \right\} , \nonumber\\
&&\left\{ r_{13},r_{23} \right\} ,\nonumber \\
&&\left\{ r_{13},r_{24} \right\} ,\nonumber \\
&&\left\{ r_{14},r_{23} \right\} ,\nonumber \\
&&\left\{ r_{14},r_{34},r_{23}=r_{24} \right\} ,\nonumber \\
&&\left\{ r_{24},r_{34},r_{12}=r_{14} \right\} ,\nonumber \\
&&\left\{ r_{24},r_{34},r_{13}=r_{14} \right\} , \\
&&\left\{ r_{24},r_{34},r_{12}=r_{13}=r_{14} \right\} ,\nonumber \\
&&\left\{ r_{13},r_{23}=r_{24} \right\} ,\nonumber \\
&&\left\{ r_{23},r_{12}=r_{13} \right\} , \nonumber\\
&&\left\{ r_{23},r_{13}=r_{14} \right\} ,\nonumber \\
&&\left\{ r_{23},r_{12}=r_{13}=r_{14} \right\} ,\nonumber \\
&&\left\{ r_{34},r_{13}=r_{14},r_{23}=r_{24} \right\} ,\nonumber \\
&&\left\{ r_{34},r_{12}=r_{13}=r_{14},r_{23}=r_{24} \right\} \nonumber .
\end{eqnarray}
Note that in all the above allowed sets, we have only brought the free
parameters and the relations that must be satisfied by them, and the zero
reaction rates have not been written. In this way we find a large class of
multi--parameter $p$--species integrable reaction--diffusion models.

\section{ two--particle conditional probabilities for 2--species model}

Now for the simplest case, that is the 2--species reactions (2), let us
calculate the two--particle conditional probabilities $P(\alpha _1,\alpha
_2,x_1,x_2;t|\beta _1,\beta _2,y_1,y_2;0)$, which is the probability of
finding particles $\alpha _1$ and $\alpha _2$ at time $t$ at sites $x_1$ and
$x_2$, respectively, if at $t=0$ we have the particles $\beta _1$ and $\beta
_2$ at sites $y_1$ and $y_2$, respectively. These probabilities can be found
by a linear combination of eigenfunctions $P(x_1,x_2)$. Therefore,
\begin{equation}
\begin{array}{l}
\left(
\begin{array}{c}
P_{AA} \\
P_{AB} \\
P_{BA} \\
P_{BB}
\end{array}
\right) (\mathbf{x;}t|\mathbf{\beta ,y};0)= \\
=\int f(p_1,p_2)e^{-\epsilon _2t}\mathbf{\Psi (}x_1,x_2)dp_1dp_2 \\
=\frac 1{(2\pi )^2}\int e^{-\epsilon _2t}e^{-i\mathbf{p.y}}\left\{ \left(
\begin{array}{c}
a \\
b \\
c \\
d
\end{array}
\right) e^{i(p_1x_1+p_2x_2)}+S_{12}(p_1,p_2)\left(
\begin{array}{c}
a \\
b \\
c \\
d
\end{array}
\right) e^{i(p_2x_1+p_1x_2)}\right\} dp_1dp_2 .
\end{array}
\end{equation}
In these expansion, $P(\mathbf{x};t|\mathbf{\beta },\mathbf{y};0)$ stands
for $P(\alpha _1,\alpha _2,x_1,x_2;t|\beta _1,\beta _2,y_1,y_2;0)$ and $%
f(p_1,p_2)$ is the coefficient of expansion, where in the second equality we
choose it to be $\frac 1{(2\pi )^2}\int e^{-i\mathbf{p.y}}$
(see \cite{r14} - \cite{r17}).
$\epsilon _2=2-e^{-ip_1}-e^{-ip_2}$ (see (17)) and $\mathbf{%
\Psi }$ is the two--particle wave function (18), in which eq.({24})
has been used for $\mathbf{A}_{\sigma _1}(\mathbf{A}_{\sigma
_1}=S_{12}(p_1,p_2)\mathbf{A}_1).$ The column matrix $\left(
\begin{array}{c}
a \\
b \\
c \\
d
\end{array}
\right) $stands for $\mathbf{A}_1$, which its components must be determined
by initial condition and $S_{12}(p_1,p_2)$ is:
\begin{equation}
S_{12}(p_1,p_2)=\left(
\begin{array}{cccc}
s_1 & 0 & 0 & 0 \\
0 & s_1 & s_2 & 0 \\
0 & 0 & s_3 & 0 \\
0 & 0 & 0 & s_1
\end{array}
\right) ,
\end{equation}
where
\begin{eqnarray}
s_1 &=&\frac{1-e^{ip_2}}{e^{ip_1}-1}, \nonumber \\
s_2 &=&\frac{r(e^{ip_2}-e^{ip_1})}{(1-e^{ip_1})\left[ 1+(r-1)e^{ip_1}\right]
},   \\
s_3 &=&\frac{(1-r)e^{ip_2}-1}{1+(r-1)e^{ip_1}}.  \nonumber
\end{eqnarray}
The matrix $S_{12}(p_1,p_2)$ is obtained from eq.(26) in which the
matrix $\mathbf{b}$ in eq.(21) (with $s=0$) has been used. By
inserting eq.(37) into eq.(36), we find:
\begin{equation}
\left(
\begin{array}{c}
P_{AA} \\
P_{AB} \\
P_{BA} \\
P_{BB}
\end{array}
\right) (\mathbf{x;}t|\mathbf{\beta ,y,}0)=\left(
\begin{array}{c}
a(F_0(t)+F_1(t)) \\
b(F_0(t)+F_1(t))+cF_2(t) \\
c(F_0(t)+F_3(t)) \\
d(F_0(t)+F_1(t))
\end{array}
\right) ,
\end{equation}
in which,
\begin{eqnarray}
F_0(t) &=&\frac 1{(2\pi )^2}\int e^{-\epsilon _2t}e^{i\mathbf{p.}(\mathbf{x}-%
\mathbf{y})}dp_1dp_2, \\
F_i(t) &=&\frac 1{(2\pi )^2}\int e^{-\epsilon _2t}e^{i(\stackrel{\sim }{%
\mathbf{p}}\mathbf{.x}-\mathbf{p}.\mathbf{y})}s_i(p_1,p_2)dp_1dp_2,\qquad
(i=1,2,3).
\end{eqnarray}
In above equations, we have suppressed the $\mathbf{x}$ and $\mathbf{y}$
dependence of $F_i$'s, for simplicity, and $\stackrel{\sim }{\mathbf{p}}%
\stackrel{}{\mathbf{=(}p_2\mathbf{,}p_1)}$. Now at $t=0$, the configuration
of the system can be one of the ($A,A$), ($A,B$), ($B,A$) or ($B,B$), where
the first particle is at site $y_1$ and the second one at $y_2$, therefore
the only acceptable behavior of $F_i(0)$ ($i=0,1,2,3$) are:
\begin{eqnarray}
F_0(0) &=&\delta _{x_1,y_1}\delta _{x_2,y_2}, \nonumber \\
F_1(0) &=&F_2(0)=F_3(0)=0.
\end{eqnarray}
$F_0(0)$ is obviously correct (see eq.(40)). For other $F_i$'s, first
we must set $p_1\rightarrow p_1+i\varepsilon $ to avoid the singularity
arising from $e^{ip_1}-1$ term in denominator of $s_1$ and $s_2$ (see
[14--17]). In this way one can show that $F_1(0)=0$. But $F_2$ and $F_3$ have
another singularity because of $1+(r-1)e^{ip_1}$
term in denominator of $s_2$ and $s_3$. One can
easily show that this singularity can be avoided only when
\begin{equation}
0\leq r<2,
\end{equation}
and for this range of interaction rates, we have $F_2(0)=F_3(0)=0.$
Therefore the validity of expansion (36) is restricted to range
(43). At $t\neq 0$, we find
\begin{eqnarray}
F_0(t) &=&e^{-2t}\frac{t^{x_1-y_1}}{(x_1-y_1)!}\frac{t^{x_2-y_2}}{(x_2-y_2)!}%
, \nonumber \\
F_1(t) &=&e^{-2t}\left[ \frac{t^{x_1-y_2+1}}{(x_1-y_2+1)!}-\frac{t^{x_1-y_2}%
}{(x_1-y_2)!}\right] \sum\limits_{k=0}^\infty \frac{t^{x_2-y_1+k}}{%
(x_2-y_1+k)!},  \\
F_2(t) &=&re^{-2t}\frac{t^{x_1-y_2}}{(x_1-y_2)!}\sum\limits_{l,k=0}^\infty
\left[ \frac 1{x_1-y_2+1}-\frac 1{x_2-y_1+k+l+1}\right] (1-r)^l\frac{%
t^{x_2-y_1+k+l+1}}{(x_2-y_1+k+l)!},  \nonumber \\
F_3(t) &=&e^{-2t}\frac{t^{x_1-y_2}}{(x_1-y_2)!}\left[ \frac{(1-r)t}{%
(x_1-y_2+1)!}-1\right] \sum\limits_{k=0}^\infty (1-r)^k\frac{t^{x_2-y_1+k}}{%
(x_2-y_1+k)!}.  \nonumber
\end{eqnarray}

One can now obtain the two--particle conditional probabilities for different
initial conditions:

1. If at $t=0$, the particles $\beta _1=\beta _2=A$ were at $y_1$ and $y_2$,
respectively, we must take $a=1$ and $b=c=d=0.$ So at $t\neq 0$, we have
\begin{equation}
P_{AA}(\mathbf{x};t|A,A,\mathbf{y};0)=F_0(t)+F_1(t),
\end{equation}
and all other $P$'s are zero.

2. If $\beta _1=A$ and $\beta _2=B$, we must take $b=1$ and $a=c=d=0.$
Therefore the only nonzero probability is
\begin{equation}
P_{AB}(\mathbf{x};t|A,B,\mathbf{y};0)=F_0(t)+F_1(t).
\end{equation}

3. If $\beta _1=B$ and $\beta _2=A$, we have $c=1$ and $a=b=d=0.$ So $%
P_{AA}=P_{BB}=0$ and
\begin{eqnarray}
P_{AB}(\mathbf{x};t|B,A,\mathbf{y};0) &=&F_2(t), \nonumber \\
P_{BA}(\mathbf{x};t|B,A,\mathbf{y};0) &=&F_0(t)+F_3(t).
\end{eqnarray}

4. And finally if $\beta _1=\beta _2=B$, we have $d=1$ and $a=b=c=0.$
So the only nonzero probability is
\begin{equation}
P_{BB}(\mathbf{x};t|B,B,\mathbf{y};0)=F_0(t)+F_1(t).
\end{equation}
Note that the appearance of the above probabilities is in agreement with our
reactions (2).

As another check of our results, it may be interesting to study the long
time behavior of these probabilities, in special $P_{AB}(\mathbf{x};t|B,A,%
\mathbf{y};0)$ which is the only nondiagonal nontrivial case. We expect that
if at $t=0$ we have a $B$ particle at site $y_1$ and an $A$ particle at site $%
y_2$ (with $y_2>y_1$), we must certainly have two
$B$ particles at $t\rightarrow \infty $ somewhere at $y_1\leq x_1<x_2$ and
$y_2\leq x_2<\infty $ sites. In other words, we expect
\begin{equation}
\sum\limits_{x_2=y_2}^\infty \sum\limits_{x_1=y_1}^{x_2-1}P_{AB}(\mathbf{x}%
;t\rightarrow \infty |B,A,\mathbf{y};0)\rightarrow 1.
\end{equation}
After some calculations, one can show that
\begin{equation}
\sum\limits_{x_2=y_2}^\infty \sum\limits_{x_1=y_1}^{x_2-1}P_{AB}(\mathbf{x}%
;t|B,A,\mathbf{y};0)=e^{-2t}\sum\limits_{n=0}^\infty \left[
1-(1-r)^{n+1}\right] \left[ I_{n+y_2-y_1}(2t)+I_{n+y_2-y_1+1}(2t)\right] ,
\end{equation}
where $I_n(x)$ is the $n$--th order Bessel function of the first kind. To
obtain the long--time behavior of (50), one may simply use the
following asymptotic form of $I_n(x)$ at $x\rightarrow \infty ,$%
\begin{equation}
I_n(x)\rightarrow \frac{e^x}{\sqrt{2\pi x}},
\end{equation}
in eq.(50). But it is not correct since eq.(51) is only valid
for $x>n$, but in eq.(50) we have a sum over $\ n$ where for every
large definite $t$, there exist infinite number of $n$ which are greater than
$t$
(it can be shown that if one calculates this limit without noting this point,
finds infinity for eq.(50), which is obviously wrong). To calculate
this limit, if one uses the identity $\sum\limits_{n=-\infty }^\infty
I_n(x)=e^x$ and takes advantage of equality $I_n(x)=I_{-n}(x)$, can show
that $\sum\limits_{n=0}^\infty I_{n+k}(x)=\frac
12 (e^x-\sum\limits_{n=-k+1}^{k-1}I_n(x)) $. Therefore
\begin{equation}
e^{-2t}\sum\limits_{n=0}^\infty \left[ 1-(1-r)^{n+1}\right]
I_{n+k}(2t)=\frac
12-e^{-2t}\sum\limits_{n=-k+1}^{k-1}I_n(2t)-e^{-2t}\sum\limits_{n=0}^\infty
(1-r)^{n+1}I_{n+k}(2t).
\end{equation}
Now in the second term of the RHS of (52) $n$ is bounded, so eq.(51)
can be used for it which leads to zero in $t\rightarrow \infty $
limit. For the third term we note that $-1<1-r<1$ (see eq.(43)), so $%
(1-r)^{n+1}\rightarrow 0$ for large $n$ . One can show that this extra $%
(1-r)^{n+1}$ factor causes the third term in RHS of (52) goes also to
zero at $t\rightarrow \infty $ limit. Therefore the $t\rightarrow \infty
\lim $it of RHS of (51) is equal to 1/2, from which eq.(49) is proved.

\section{ 2--species model with left--right diffusion}

In this section we want to study the range of parameters (reaction rates),
which makes the reactions (4) integrable. In this case the master
equation is
\begin{eqnarray}
\frac \partial {\partial t}P_{\alpha _1...\alpha _N}(x_1,...,x_N;t)
&=&D_R\sum\limits_{i=1}^NP_{\alpha _1...\alpha
_N}(x_1,...,x_{i-1},x_i-1,x_{i+1},...,x_N;t)\nonumber \\
&&+D_L\sum\limits_{i=1}^NP_{\alpha _1...\alpha
_N}(x_1,...,x_{i-1},x_i+1,x_{i+1},...,x_N;t)  \nonumber \\
&&-NP_{\alpha _1...\alpha _N}(x_1,...,x_N;t),
\end{eqnarray}
in which we have used a time scale so that
\begin{equation}
D_R+D_L\equiv 1.
\end{equation}
The boundary conditions are
\begin{eqnarray}
D_RP_{AB}(x,x)+D_LP_{AB}(x+1,x+1) &=&rP_{BA}(x,x+1)+(1-s)P_{AB}(x,x+1), \nonumber \\
D_RP_{BA}(x,x)+D_LP_{BA}(x+1,x+1) &=&sP_{AB}(x,x+1)+(1-r)P_{BA}(x,x+1), \\
D_RP_{\alpha \alpha }(x,x)+D_LP_{\alpha \alpha }(x+1,x+1) &=&P_{\alpha \alpha }(x,x+1),
\qquad (\alpha =A,B).  \nonumber
\end{eqnarray}
Note that for $D_L=0,$ eqs.(53) and (55) reduce to eqs.(5)
and (6), respectively, and for $r=s=0$, these equations lead to
corresponding ones in \cite{r14}.

We must proceed the same steps as previous section, where in this case lead
to the following relations for energy and boundary conditions:
\begin{equation}
\epsilon _N=\sum\limits_{j=1}^N(1-D_Re^{-ip_j}-D_Le^{ip_j}),
\end{equation}
and
\begin{equation}
D_R\mathbf{\Psi (...,}\zeta ,\zeta ,...)+D_L\mathbf{\Psi (...,}\zeta
+1,\zeta +1,...)=\mathbf{b}_{k,k+1}\mathbf{\Psi (...,}\zeta ,\zeta
+1,...),
\end{equation}
where $\mathbf{b}_{k,k+1}$ is defined through (20) with $\mathbf{b}$ in
eq.(21). If we substitute the coordinate Bethe ansatz (18) in
eq.(57), the relation between coefficients is like eq.(24), but
now with following S-matrix:
\begin{equation}
\mathbf{S(}z_1,z_2)=-(D_R+z_1z_2D_L\mathbf{-}z_{1\mathbf{\ }}\mathbf{b}%
)^{-1}(D_R+z_1z_2D_L\mathbf{-}z_{2\mathbf{\ }}\mathbf{b}).
\end{equation}
This S--matrix must satisfy the QYBE (28). Like the previous case, here
there are also fourteen equations that must be solved for $r,s$ and an extra $%
D_R$ parameters ($D_L$ is fixed by $D_R$ through eq.(54)). These
equations are highly nonlinear and we are not able to solve them exactly,
even by using the standard programs like MAPLE. So we restrict
ourselves to the cases in which
\begin{equation}
r+s=1.
\end{equation}
In this way, we can find the complete set of solutions. We believe that there
are no other solutions even if the constraint (59) is removed (we
have checked many other cases, but no one satisfied QYBE).
The solutions with $r=0$ or $s=0$ are not new. They are the known
models like the models introduced in (29), or the simple diffusion
models introduced in \cite{r14} (in which both $r$ and $s$ are zero). There
are only two new integrable models, as follows:
\begin{eqnarray}
&&A+\emptyset \stackrel{D_R}{\rightarrow }\emptyset +A,\nonumber \\
&&B+\emptyset \stackrel{D_R}{\rightarrow }\emptyset +B,  \nonumber \\
&&\emptyset \stackrel{D_L}{+A\rightarrow A+}\emptyset ,   \\
&&\emptyset \stackrel{D_L}{+B\rightarrow B+}\emptyset ,  \nonumber \\
&&B+A\stackrel{D_R}{\rightarrow }A+B,  \nonumber \\
&&A+B\stackrel{D_L}{\rightarrow }B+A,  \nonumber
\end{eqnarray}
and
\begin{eqnarray}
&&A+\emptyset \stackrel{D_R}{\rightarrow }\emptyset +A,\nonumber \\
&&B+\emptyset \stackrel{D_R}{\rightarrow }\emptyset +B,  \nonumber \\
&&\emptyset \stackrel{D_L}{+A\rightarrow A+}\emptyset ,  \\
&&\emptyset \stackrel{D_L}{+B\rightarrow B+}\emptyset ,  \nonumber \\
&&B+A\stackrel{D_L}{\rightarrow }A+B,  \nonumber \\
&&A+B\stackrel{D_R}{\rightarrow }B+A,  \nonumber
\end{eqnarray}
in which $D_R=1-D_L$. In
this way we find two one--parameter family integrable models, which their time
dependent probabilities can be found by eq.(9), with $\epsilon _N$ in
eq.(56), $\mathbf{\Psi }$ is given by (18) and (24), with
S--matrix introduced in (58), and $\mathbf{b}$ in (21).

\section{Acknowledgment}

We would like to thank the research council of the University of Tehran for
partial financial support.

\end{document}